\documentclass[twocolumn,prx,superscriptaddress]{revtex4-1}

\usepackage{graphicx}
\usepackage{dcolumn}
\usepackage{bm}
\usepackage{url}
\usepackage{sistyle}

\begin{document}

\title{Anisotropic Etching of Graphite and Graphene in a Remote Hydrogen Plasma} 

\author{D.~Hug}
\affiliation{Department of Physics, University of Basel, CH-4056 Basel, Switzerland}

\author{S.~Zihlmann}
\affiliation{Department of Physics, University of Basel, CH-4056 Basel, Switzerland}

\author{M. K.~Rehmann}
\affiliation{Department of Physics, University of Basel, CH-4056 Basel, Switzerland}

\author{Y. B. Kalyoncu}
\affiliation{Department of Physics, University of Basel, CH-4056 Basel, Switzerland}

\author{T. N. Camenzind}
\affiliation{Department of Physics, University of Basel, CH-4056 Basel, Switzerland}

\author{L.~Marot}
\affiliation{Department of Physics, University of Basel, CH-4056 Basel, Switzerland}

\author{K. Watanabe}
\affiliation{National Institute for Material Science, 1-1 Namiki, Tsukuba 305-0044, Japan}

\author{T. Taniguchi}
\affiliation{National Institute for Material Science, 1-1 Namiki, Tsukuba 305-0044, Japan}

\author{D.~M.~Zumb\"uhl}
\email[]{dominik.zumbuhl@unibas.ch}
\affiliation{Department of Physics, University of Basel, CH-4056
Basel, Switzerland}


\date{\today}

\begin{abstract}
We investigate the etching of a pure hydrogen plasma on graphite samples and graphene flakes on SiO$_2$ and hexagonal Boron-Nitride (hBN) substrates. The pressure and distance dependence of the graphite exposure experiments reveals the existence of two distinct plasma regimes: the {\em direct} and the {\em remote} plasma regime. Graphite surfaces exposed {\em directly} to the hydrogen plasma exhibit numerous etch pits of various size and depth, indicating continuous defect creation throughout the etching process. In contrast, anisotropic etching forming regular and symmetric hexagons starting only from preexisting defects and edges is seen in the {\em remote} plasma regime, where the sample is located downstream, outside of the glowing plasma. This regime is possible in a narrow window of parameters where essentially all ions have already recombined, yet a flux of H-radicals performing anisotropic etching is still present. At the required process pressures, the radicals can recombine only on surfaces, not in the gas itself. Thus, the tube material needs to exhibit a sufficiently low H radical recombination coefficient, such a found for quartz or pyrex. In the {\em remote} regime, we investigate the etching of single layer and bilayer graphene on SiO$_2$ and hBN substrates. We find {\em isotropic} etching for single layer graphene on SiO$_2$, whereas we observe highly {\em anisotropic} etching for graphene on a hBN substrate. For bilayer graphene, anisotropic etching is observed on both substrates. Finally, we demonstrate the use of artificial defects to create well defined graphene nanostructures with clean crystallographic edges.
\end{abstract}

\maketitle
Graphene nanoribbons (GNRs) have emerged as a promising platform for graphene nano devices, including a range of intriguing quantum phenomena beyond opening of a confinement induced band gap\cite{Fujita,Nakada,Son_2006,Son,Trauzettel}. In armchair GNRs, giant Rashba spin-orbit coupling can be induced with nanomagnets, leading to helical modes and spin filtering\cite{Klinovaja}. Further, Majorana fermions localized at the ends of the ribbon were predicted in proximity of an s-wave superconductor\cite{Klinovaja}. Zigzag ribbons, on the other hand, were proposed as a promising system for spin filters\cite{Son_2006}. Theory showed that electronic states in zigzag ribbons are strongly confined to the edge\cite{Fujita,Nakada,Son_2006}, recently observed in experiments\cite{Tao,Pan_2012,Zhang_2013,wang2016giant}. Further, edge magnetism was predicted to emerge at low temperatures\cite{Fujita,Nakada,Son,YazyevRPP}, with opposite GNR edges magnetized in opposite directions. High quality, crystallographic edges are very important here, since edge disorder suppresses magnetic correlations\cite{YazyevRPP} and tends to cause electron localization, inhibiting transport studies. GNRs fabricated with standard electron beam lithography (EBL) and Ar/O$_2$ etching typically exhibit pronounced disorder \cite{Han_2007,MuccioloPRB79,OostingaPRB81,StampferPRL,GallagherPRB81_2010,LiuPRB80,MolitorSST}, complicating transport studies.

Fabrication methods creating ribbons with clean crystallographic edges were recently developed, including carbon nanotube unzipping \cite{Jiao,Kosynkin_2009}, ultrasonication of intercalated graphite \cite{Li}, chemical bottom up approaches \cite{Cai,ruffieux2016surface}, anisotropic etching by nickel nanoparticles \cite{campos2009anisotropic} or carbothermal etching of graphene sheets \cite{Nemes,Krauss}. Here, we use a hydrogen (H) plasma etching technique \cite{McCarroll,Yang,Shi,Xie} because it allows precise, top-down and on-demand positioning and tailoring of graphene nanostructures. Such nanostructures can easily be designed to spread out into larger graphene areas incorporated into the same graphene sheet, thus providing for a relatively easy way to make electrical contacts.

In this work, we investigate the anisotropic H plasma etching of graphite surfaces in dependence of the gas pressure and the sample - plasma distance. We find that the etching characteristics can be divided into a direct and a remote plasma regime. In the {\em direct} plasma regime, the sample is placed within the glowing plasma, and surfaces show many hexagons of various sizes indicating a continuous defect induction throughout the etching process. In the {\em remote} plasma regime, on the other hand, the sample is placed downstream of the glowing plasma, and etching occurs only from preexisting defects which makes the fabrication of well defined graphene nanostructures possible. Further, we have prepared single layer (SL) and bilayer (BL) graphene flakes on SiO$_2$ and hexagonal boron nitride (hBN) substrates and exposed them to the remote H plasma. We observe a strong dependence of the anisotropy of the etch on the substrate material. SL graphene on SiO$_2$ is etched isotropically, confirming previous findings\cite{Shi,Diankov}, whereas we observe highly anisotropic etching of SL graphene on hBN \cite{wang2016patterning}, producing very regular and symmetric hexagonal etch pits. Anisotropic etching of SL graphene on hBN offers the possibility to fabricate diverse graphene nanostructure with well defined edges (e.g. GNRs) and allows investigation of their intrinsic electronic transport properties.

A pure H plasma was created in a quartz tube through a matching network by a $13.56\,$MHz radio frequency (RF) generator at a typical power of $30\,$W. This RF power was capacitively coupled to the $80\,$mm diameter tube by an outer electrode acting as a surfatron\cite{Moisan}. The pressure was regulated using a needle valve for 20 SCCM H gas flow of purity 6N. The sample was placed at a distance $d$ from the end of the surfatron, was electrically floating and a three-zone furnace controlled the temperature $T$. See supplementary online materials (SOM) for additional information. Ion impact energy is roughly the difference between the plasma potential and the floating potential and is around $10-15$ eV with an average ion mass of 2 amu. We estimate the ion flux to be significantly lower than $10^{15}$ ions/cm$^{2}$s measured for a similar plasma setup but at lower pressure\cite{Eren}. In order to characterize and optimize the anisotropic etching process, we studied the influence of pressure, distance, and temperature on the etching process, generally finding good repeatability. We first investigated graphite flakes, allowing for rather simple and fast processing. The graphite specimen\cite{NGS} were cleaned by peeling with scotch tape and subsequently exposed for one hour to a pure H plasma at $T=400\,\degC$.

\begin{figure}[t]\hspace*{2mm}
\includegraphics[width=8.5cm]{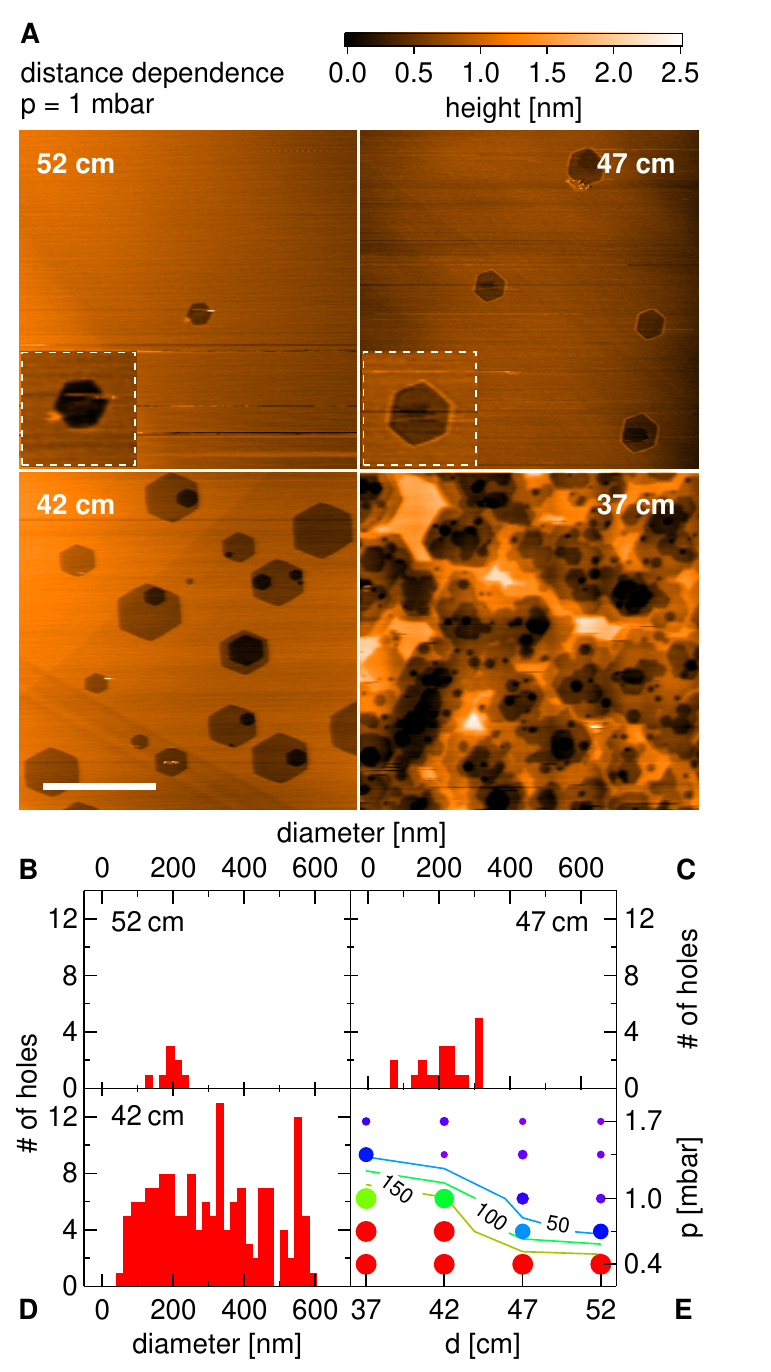}
\vspace{-6mm}\caption{\label{fig:figure1}{\bf Distance dependence of graphite exposures}\newline (A) AFM images (tapping mode) of graphite surfaces for various distances $d$, as labeled, all exposed to the plasma for $1\,$h at $p=1\,$mbar and $T=400^\circ$C, all shown on the same color scale. Main panels are $3\times3\,\mathrm{\mu m^{2}}$, scale bar is $1\,\mathrm{\mu m}$, insets (dashed white boxes) are $0.25\times0.25\,\mathrm{\mu m^2}$. Slight hexagon distortion at $42\,$cm is an imaging artefact due to drift. (B-D) Histograms obtained from $10\times10\,\mathrm{\mu m^{2}}$ scans, showing the number of holes against hole diameter (bin size $20\,$nm). (E) The size of the circle markers corresponds to the width of the diameter distribution. The color indicates the number of holes, with red corresponding to large number of holes. For samples located within the glowing plasma (red circles), a lower bound of $300$ holes and a minimum width of distribution of diameter of $600\,$nm is shown.} \vspace{-7mm}
\end{figure}

We first present the distance dependence of the H plasma process. Figure\,\ref{fig:figure1}A shows AFM topography scans for exposures of one hour at four different distances at constant pressure $p=1\,$mbar. At the larger distances, etch pits of monolayer step height are created upon plasma exposure, exhibiting a regular hexagonal shape and demonstrating a strongly anisotropic process \cite{McCarroll,Yang}. All observed hexagons exhibit the same orientation. From previous studies, it is known that hexagons created by exposure to a remote H plasma exhibit edges pointing along the zigzag direction \cite{McCarroll,Yang}. As the sample is brought closer to the plasma, significantly more etch pits appear, often located at the border of existing holes, sharing one common hexagon side (see Figure\,\ref{fig:figure1}A, $d=42\,$cm). For the closest position $d=37\,$cm -- unlike the larger distances -- the sample is located within the visible plasma glow region, resulting in a strong and several layers deep scarring of the entire surface.

To quantitatively study the distance dependence, we evaluated larger images to gather better statistics and plot histograms showing the number of holes as a function of diameter, see Figure\,\ref{fig:figure1}B-D. The overall number of holes obviously increases strongly with decreasing sample-surfatron distance $d$. For small distances, a wide distribution of diameters is seen, ranging from several $100\,$nm down to nearly vanishing hexagon size, suggesting that new defects serving as etch seeds are created throughout the exposure time. The width of the hole diameter distribution is given by the anisotropic etch rate and the exposure duration in this regime. For larger $d$, on the other hand, the few holes seen have comparable diameters, consistent with etching proceeding predominantly from preexisting graphite defects, without adding new defects. This results in a narrow width of the distribution of hole sizes. As previously reported \cite{McCarroll,Shi,Yang}, exposure to energetic ions seems to create defects, while exposure to hydrogen radicals appears to result in anisotropic etching and growth of hexagons centered around preexisting defects and borders.

\begin{figure}[t]\hspace*{2mm}
\includegraphics[width=8.5cm]{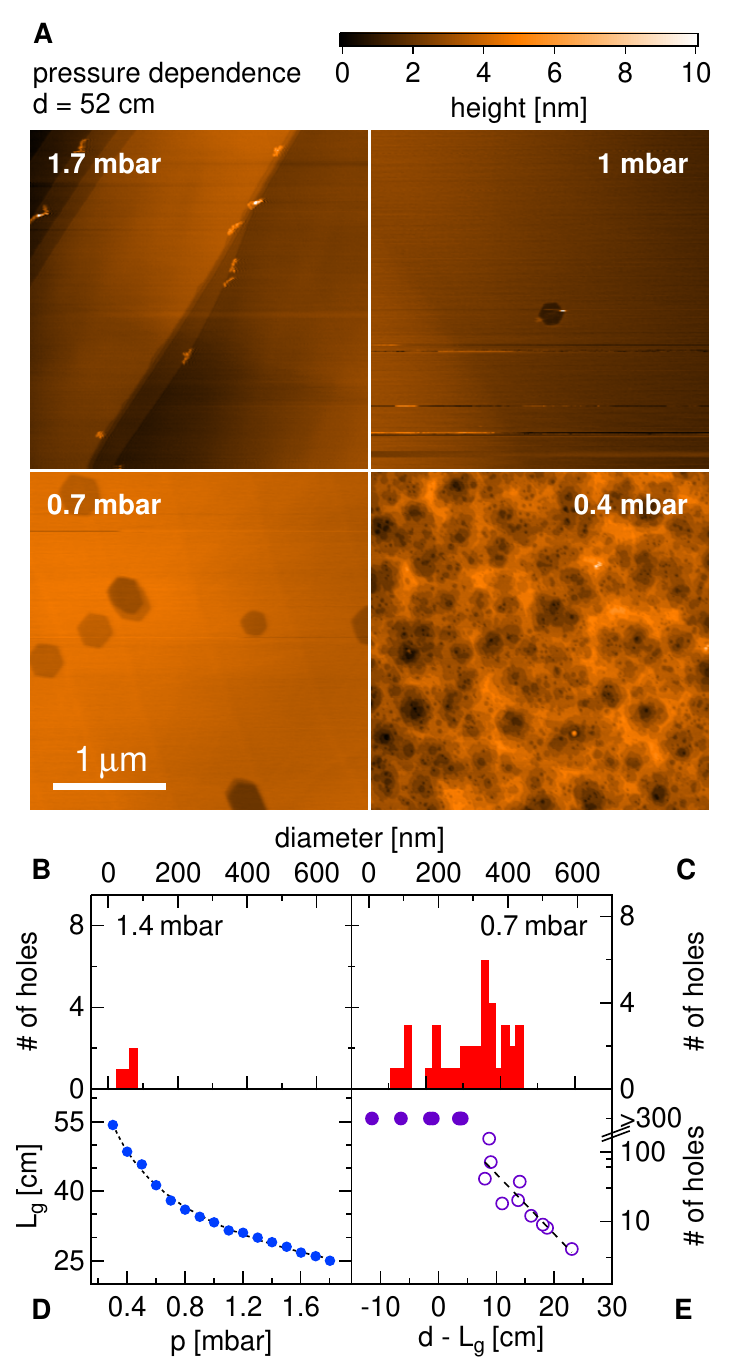}
\vspace{-7mm}\caption{\label{fig:figure2} {\bf Pressure dependence of graphite exposures}\newline (A) AFM images (tapping mode) of graphite surfaces for various $p$, as indicated, exposed for one hour at $d=52\,$cm and $T=400^\circ$C, all shown on the same color scale. All panels are $3\times3\,\mathrm{\mu m^2}$. (B,C) Histograms from $10\times10\,\mathrm{\mu m^{2}}$ scans, displaying the number of holes against hole diameter (bin size $20\,\mathrm{nm}$) for $p$ as labeled. (D) Length $L_g$ of the optically visible plasma as a function of $p$. The dashed curve is a $1/\sqrt{p}$ fit. (E) Number of holes versus distance from plasma edge $d-L_g$. A lower bound of $300$ holes is given for the heavily etched cases where an exact hole-count was not feasible. The dashed black line is an exponential fit to the data with $<300$ holes with 1/e decay length $\sim5\ $cm. }\vspace{-7mm}
\end{figure}

Next, we turn to the pressure dependence. In Figure\,\ref{fig:figure2}A, AFM topography images are shown at four different pressures $p$ at constant distance $d=52\,$cm. The number of holes increases with decreasing pressure, similar to decreasing distance, giving rise to etch pits of monolayer step height at intermediate pressures. At the highest pressures, however, no etch pits were observed, in strong contrast to the lowest pressure, where ubiquitous and deep etching is seen, demonstrating the strong influence of $p$. Analyzing the etch pits using histograms confirms that $p$ and $d$ have a similar influence on the etching process (compare Figure\,\ref{fig:figure2}B, C with Figure\,\ref{fig:figure1}B-D). Figure\,\ref{fig:figure1}E summarizes the histograms of all investigated graphite samples, using color to represent the number of holes, while the size of each marker is proportional to the width of the distribution of hole diameters. A clear correlation between the number of holes and the width of the distribution is seen: the largest circles are red, while the small circles are purple.

The analysis of the graphite exposure data leads to two qualitatively different types of processes: the {\em direct} and the {\em remote} plasma regime. In the {\em direct} plasma regime (large, red circles, Figure\,\ref{fig:figure1}E), the sample is located directly within the plasma discharge region, hence exposing it to large densities of radicals and ions, capable of inducing defects throughout the exposure, giving a broad hole diameter distribution. In the {\em remote} plasma regime (small, purple circles, Figure\,\ref{fig:figure1}E), on the other hand, the sample is positioned outside, downstream of the plasma generation region, where ions have recombined and only a residual flux of radicals is present. There, etching proceeds predominantly from preexisting defects and edges, leaving the basal planes mostly untouched. In this regime, a narrow distribution of hole diameters results, centered around the diameter given by the anisotropic etch rate and the exposure time.

Further, there is an intimate connection between distance and pressure: lower pressure results in a longer gas mean free path and therefore a larger average distance for recombination in the diffusive gas. This results in a larger length of the plasma column $L_g(p)$, measured from the edge of the visibly glowing plasma to the surfatron, see Figure\,\ref{fig:figure2}D. Thus, changing the pressure with fixed sample position modifies the distance between sample and plasma edge. Hence, it is useful to introduce an effective distance $d^{\prime}=d-L_g(p)$, the distance from the sample to the edge of the glowing plasma. Thus, $d^\prime\lesssim0$ roughly marks the direct plasma regime while $d^\prime\gg0$ signifies the remote plasma regime. Reactive particles are generated inside the plasma column and start recombining once they have left the plasma generation region.

The reaction kinetics in low temperature H plasmas are highly non-trivial despite the relatively simple chemical composition\cite{SammBook}. Nevertheless, it is well known that at the pressures used here ($p\sim1\,$mbar), the predominant radical decay mechanism is surface mediated association rather than gas collisions. Two colliding H atoms require a third body to carry away the excess energy for association to occur \cite{DixonLewis1962}. However, under the present conditions, three body collisions are very unlikely, thus leaving only the surface assisted process (which also leads to surface heating\cite{Grubbs}). Recombination of ions, in contrast, can also occur through an additional collisional channel, in absence of a surface. Which species -- ions or radicals -- decay on a shorter length scale downstream of the plasma edge thus depends on both the surface properties and gas parameters. For anisotropic etching without defect creation, a flux of H radicals in absence of ions is needed, thus requiring the ion density to decay on a shorter length than the radicals.

The surface attenuation of H radicals thus plays an important role, and was previously studied \cite{Shuler1949,Grubbs}. Some glasses such as pyrex or quartz -- as used in our experiments -- were identified as a materials with a low recombination coefficient, particularly compared to some common metallic surfaces such as stainless steel and aluminum. This weak surface attenuation can open a downstream window offering a flux of H radicals while essentially all ions have already recombined, as desired and achieved here, see e.g. Figure\,\ref{fig:figure1}B, \ref{fig:figure2}B and 3 (below). Nevertheless, the etch rate in the downstream window was observed to decrease slowly over long periods of time, reaching a vanishingly small etch rate after more than 100 hours of plasma exposure. The elevated temperatures in the furnace may enhance impurity migration towards the surfaces of the tube, possibly amplifying the surface attenuation of H radicals. Larger anisotropic etch rates were observed when utilizing higher purity quartz tubes manufactured from synthetic fused silica\cite{Suprasil}, supporting the assumption of the role of impurities. High impurity content and even small amounts of metallic deposition on the tube wall give wave damping due to dielectric losses and result in an enhanced decay of radicals.

To study the decay of reactive species, we note that the ion flux is proportional to the number of holes created. We find a roughly exponential decrease of the number of holes with distance, see Fig.\ref{fig:figure2}E and SOM, with a 1/e decay length of about 5 cm. The anisotropic etch rate, on the other hand, is related to the flux of H radicals. We extract the anisotropic etch rate, defined as the growth per unit time of the radius of a circle inscribed to the hexagonal etch pit, averaged over a number of  holes, shown in Figure\,\ref{fig:figure4}A. Only the largest set of hexagons of each exposed graphite sample were evaluated to obtain the etch rate, since smaller holes might not have etched from the beginning of the exposure. As expected, the anisotropic etch rate is largest for small distances, falling off quickly with increasing separation from the plasma edge. There is also an apparent pressure dependence, with larger pressures tending to give lower etch rates, see Fig.~\ref{fig:figure4}A. Given only two or three points along the d-axis for each pressure, and only few holes for some parameter sets ($d$, $p$), a reliable H-radical decay length cannot be extracted from these data. A theoretical estimate gives an H-radical decay length of $\sim12\,$cm, see SOM, in agreement with observations in Fig.~\ref{fig:figure4}A, and longer than the ion decay length of $5\,$cm, as observed. The etch rates we extract are a few nm per min at $400\,{\rm ^\circ C}$, consistent with previous reports\cite{Yang,Shi}.

\begin{figure}[t]\hspace*{-3mm}
\includegraphics[width=7.5cm]{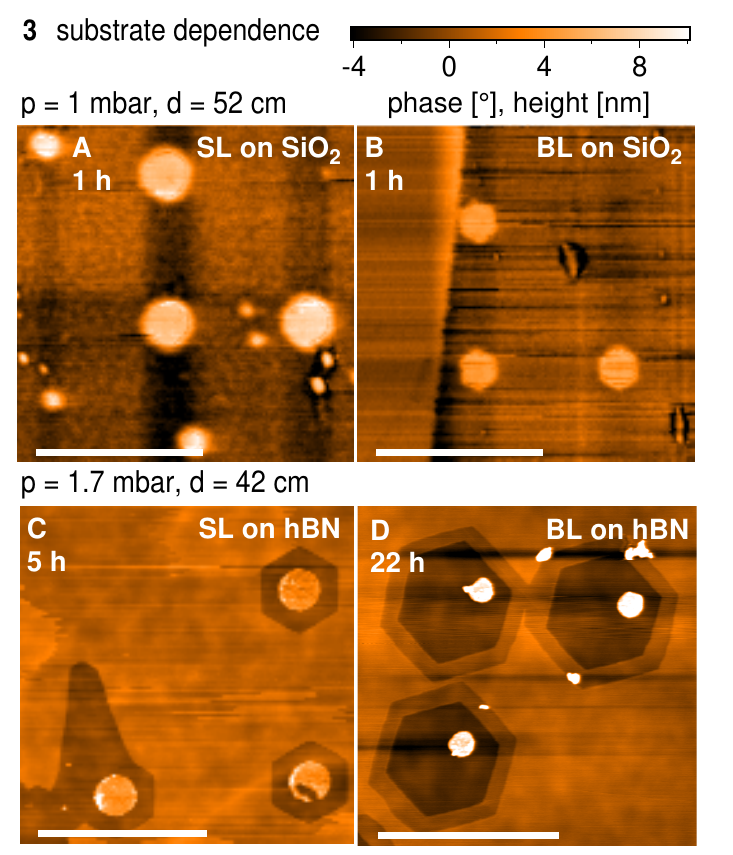}
\vspace{-2mm}\caption{\label{fig:figure3}{\bf Substrate dependence of SL/BL graphene} (A,B) AFM phase contrast images of a SL (A) and BL (B) section of the same flake on a SiO$_2$ substrate, etched for $1\,$h at $T=450\,\degC$. Round holes of $50\,$nm diameter were defined before H-etching. AFM topography image of a SL (C) and BL (D) flake on hBN etched for $5\,$h and $22\,$h, respectively. Holes of $200\,$nm (SL) and $100\,$nm (BL) were defined before etching. For (D) the color scale values are divided by four. The scale bars on all images are $1\,\mathrm{\mu m}$.}\vspace{-5mm}
\end{figure}

Next, we study the plasma exposure of SL and BL graphene exfoliated onto a SiO$_2$ substrate using the established tape method\cite{novoselov2004electric}. We patterned disks using standard EBL and reactive ion etching with an $\mathrm{Ar/O}_2$ plasma, resulting in circular graphene holes which were subsequently exposed to the remote H plasma in the regime where H radicals but essentially no ions are present, as determined from the graphite experiments. BL graphene grows regular hexagons with parallel sides (see Figure\,\ref{fig:figure3}B), as expected from the graphite results. SL graphene, on the other hand, displayed mostly round holes (see Figure\,\ref{fig:figure3}A), though some weakly developed, irregular hexagonal shapes are also occasionally seen. Further, several additional, not EBL defined holes appear on the SL after exposure, all smaller than the EBL initiated etch pits. After a second plasma exposure, the number of holes on the SL increased further, indicating generation of new defects, while only EBL defined holes appear on the BL. Note that the SL and BL regions shown in Figure\,\ref{fig:figure3}A and B are located on the same graphene flake, ensuring identical plasma conditions.

In addition, the average hole diameter on SL is visibly larger than on the BL (Figure\,\ref{fig:figure3}A and B) after the same exposure time, indicating a faster etch rate on SL. Thus, SL on SiO$_2$ is more reactive when exposed to the plasma and no longer anisotropic when exposing . This is consistent with previous reports\cite{Diankov,Shi,wang2016patterning}, and is suspected to arise from charge inhomogeneities in the SiO$_2$ substrate\cite{Zhang,DeckerNL2011,XueNM2011} or other SiO$_2$ surface properties. A broad range of plasma parameters in the remote regime were investigated for SL and BL samples on SiO$_2$, giving qualitatively similar results (isotropic SL etching). The etch rate for SL and BL on SiO$_2$ is shown in Figure\,\ref{fig:figure4}B. For the SL samples, only the EBL defined holes were evaluated, ignoring the plasma induced defects, since these do not etch from the beginning of the exposure. Clearly, for all plasma parameters studied, SL exhibits a significantly larger etch rate compared to BL \cite{Diankov,Yang}, as already visible from the AFM images in Figure\,\ref{fig:figure3}A and B. The temperature dependence of the etch rate for both SL and BL on SiO$_2$ is shown in Figure\,\ref{fig:figure4}C. The etch rates are strongly reduced at temperatures far above and below the process temperature, consistent with previous reports \cite{Yang,Diankov}, and consistent with reported hydrogen recombination rates on quartz increasing dramatically with temperature \cite{Kim}.

\begin{figure}[t]\hspace*{-3mm}
\includegraphics[width=7.5cm]{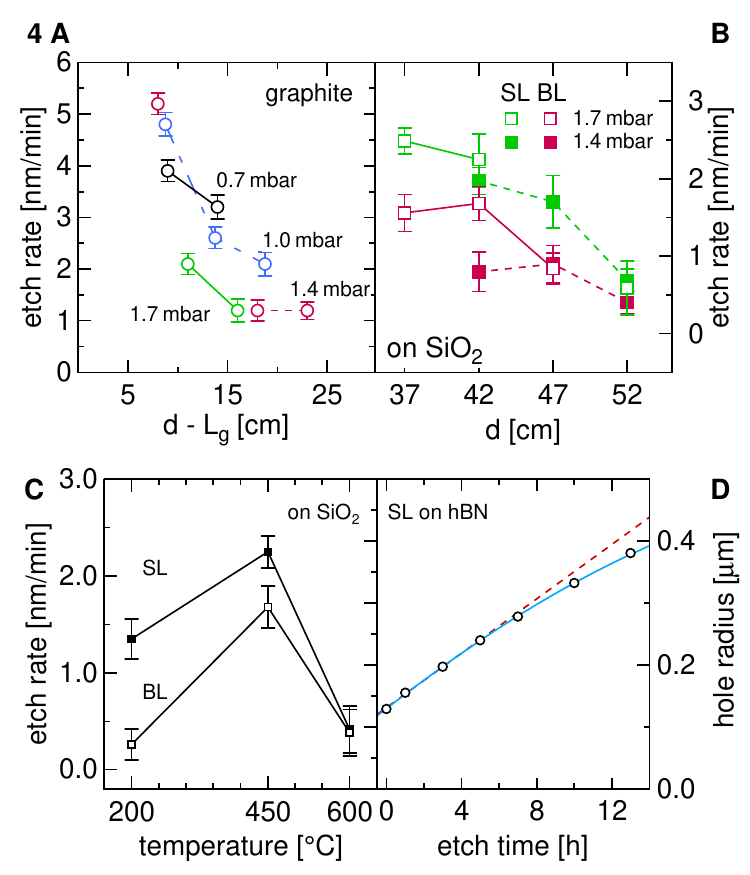}
\vspace{-3mm}\caption{\label{fig:figure4} {\bf Anisotropic etch rates} (A) Graphite anisotropic etch rate versus distance from plasma $d-L_g$ for several configurations. (B) Etch rate of SL and BL on SiO$_2$ at indicated parameters. (C) Temperature dependence of the etch rate of SL and BL samples on SiO$_2$. (D) Average radius of a circle inscribed to the hexagonal etch pits as a function of exposure time for SL on hBN. Several etch pits were evaluated in order to obtain average size and standard deviation, where the latter is smaller than the diameter of the marker circle. The dashed red line is a linear fit to the points at $\leq 5\,$h, the blue curve is a tanh-fit shown as a guide for the eye. }\vspace{-5mm}
\end{figure}

To study the substrate dependence, we investigate SL and BL graphene on high-quality hBN crystals, as in Ref.\,\onlinecite{Taniguchi}. SL and BL graphene were aligned and deposited onto areas covered with several $10\,$nm thick hBN lying on a SiO$_2$ substrate, following the recipe of Ref.\,\onlinecite{Dean}. Then, the same fabrication steps were repeated as before to fabricate circular graphene holes. Figure\,\ref{fig:figure3}C shows an AFM topography image of SL graphene on hBN after $5\,$h of remote H plasma exposure. Clearly, very regular and well aligned hexagonal holes are visible, indicating a highly anisotropic etch. We observed this anisotropic SL graphene etching on hBN in more than 10 samples demonstrating the high reproducability of the process.

In Figure\,\ref{fig:figure3}D we present an AFM topography image of a BL graphene flake on hBN which was exposed to the H plasma for $22\,$h. We observe anisotropic etching of the BL flake with a slightly higher etch rate for the top layer ($\sim 0.3\,$nm/min) compared to the bottom layer ($\sim 0.2\,$nm/min), leading to a staircase-like structure at the etch pit borders. As seen in Figure\,\ref{fig:figure3}D, the hexagons in the bottom and the top layer are of the same orientation. We note that the bottom layer is on hBN while the top layer is laying on graphene. The situation of the top layer is comparable to the SL etching on a graphite surface, where it was shown that the edges of the hexagons are aligned with the zigzag direction of the graphite lattice \cite{McCarroll,Yang}. Since the bottom layer exhibits hexagons oriented in the same direction as the hexagons emerging on the top layer, this further confirms that the etching of SL graphene on hBN is yielding etch pits oriented along the zigzag direction. The ribbon defined by the two left hexagons in Figure\,\ref{fig:figure3}D has a width of about $20\,$nm, demonstrating the fabrication of nanoscale graphene structures with a remote H plasma.

The size of the SL hexagons as a function of exposure time is shown in Figure\,\ref{fig:figure4}D. A linear fit (dashed red) is clearly over estimating the etch rate for long exposure times, deviating from the data by several standard deviations for the longest times. This hints towards either an insufficient H atom collection mechanism as the etch pits are growing larger or an aging effect of the tube as discussed above.

Raman spectroscopy on SL and BL samples on hBN was performed before and after H plasma etching. The $D$ and $D^\prime$ disorder peaks were not seen (see SOM), both before and after H plasma etching. This suggests that neither defect formation nor hydrogenation\cite{Ferrari,Elias,Sofo,Eren} is occurring in the bulk 2D during plasma etching, taking into account the annealing of the sample during the cool down phase\cite{Elias}, opening the door for high quality electrical properties.

The EBL defined circles stand very clearly visible in the center of the hexagons as an elevated region, as seen in Figure\,\ref{fig:figure3}C and D, growing in height but not diameter upon further H plasma exposure. These discs appear also away from the graphene flakes directly on the hBN, wherever circles were EBL/$\mathrm{Ar/O}_2$-plasma defined. However, these elevated regions are also observed to shrink in height in ambient conditions. For a better understanding of the composition and behaviour of these surface structures, further investigations are required, which are however beyond the scope of this work. In addition, the adhesion between graphene and hBN often appears to be rather poor. Graphene flakes of several micrometres in length seem to be tilted with respect to the circular pillars induced by EBL. AFM tip forces or elevated temperatures may have shifted the flakes from their original position\cite{woods2016macroscopic,wang2016thermally}.

In conclusion, we have investigated the pressure and distance dependence of the anisotropic etching of graphite surfaces in a H plasma. We have found that the etching characteristics can be divided into two regimes, the remote and the direct plasma regime. In the remote region of the plasma ($d^\prime>0$) etching only occurs at preexisting defect sites whereas for $d^\prime<0$ new defects are induced. Further, we have prepared SL and BL graphene flakes on SiO$_2$ and hBN substrates and exposed them to the remote H plasma. We observed isotropic etching of SL graphene on SiO$_2$, whereas on hBN it is highly anisotropic, exhibiting very regular and symmetric hexagonal etch pits. BL graphene, on the other hand, did not show a substrate dependence of the etching character and was anisotropic for both substrates.

By inducing artificial defects by lithographic means it becomes possible to pattern graphene nanostructures of various geometries with clean crystallographic edges defined by the etching in a remote H plasma. This leads to the opportunity to fabricate GNRs with well defined edges on a well suited substrate for electronic transport experiments, such as hBN. It would be interesting to study the etching process in dependence of the graphene electrochemical potential, which can be adjusted in-situ with a back gate during the etching process. Also, a remote nitrogen plasma\cite{ZhangN} could be investigated to be potentially used in a similar way to define armchair edges via anisotropic etching of atomic nitrogen.

We would like to thank B.~Eren, R.~Maurand, C.~Sch\"onenberger and R.~Steiner for helpful discussions and we acknowledge supported from the Swiss Nanoscience Institute (SNI), NCCR QSIT and Swiss NSF. Growth of hexagonal boron nitride crystals was supported by the Elemental Strategy Initiative conducted by the MEXT, Japan and JSPS KAKENHI Grant Numbers JP26248061, JP15K21722 and JP25106006.
\subsection{REFERENCES}

%


\end{document}